\documentclass{jfm}

\usepackage{graphicx}
\usepackage{natbib}

\ifCUPmtlplainloaded \else
  \checkfont{eurm10}
  \iffontfound
    \IfFileExists{upmath.sty}
      {\typeout{^^JFound AMS Euler Roman fonts on the system,
                   using the 'upmath' package.^^J}%
       \usepackage{upmath}}
      {\typeout{^^JFound AMS Euler Roman fonts on the system, but you
                   dont seem to have the}%
       \typeout{'upmath' package installed. JFM.cls can take advantage
                 of these fonts,^^Jif you use 'upmath' package.^^J}%
      }
  \else
  \fi
\fi

\ifCUPmtlplainloaded \else
  \checkfont{msam10}
  \iffontfound
    \IfFileExists{amssymb.sty}
      {\typeout{^^JFound AMS Symbol fonts on the system, using the
                'amssymb' package.^^J}%
       \usepackage{amssymb}%
         \let\leq=\leqslant
         \let\geq=\geqslant
      }{}
  \fi
\fi

\ifCUPmtlplainloaded \else
  \IfFileExists{amsbsy.sty}
    {\typeout{^^JFound the 'amsbsy' package on the system, using it.^^J}%
     \usepackage{amsbsy}}
    {\providecommand\boldsymbol[1]{\mbox{\boldmath $##1$}}}
\fi

\providecommand\n{\boldsymbol{\nabla}}
\providecommand\bcdot{\boldsymbol{\cdot}}

\renewcommand\Re{\mbox{\textit{Re}}} 
\newcommand\Ra{\mbox{\textit{Ra}}}   
\renewcommand\Pr{\mbox{\textit{Pr}}} 
\newcommand\Fr{\mbox{\textit{Fr}}}   
\newcommand\Gr{\mbox{\textit{Gr}}}   

\newsavebox{\astrutbox}
\sbox{\astrutbox}{\rule[-5pt]{0pt}{20pt}}

\newcommand\p{\ensuremath{\partial}}

\def \f#1{\ensuremath{\boldsymbol{#1}}}
\def \n   {\vec{\nabla}\!}
\def \p   {\partial}

\def \Om  {\Omega}
\def \heta{\eta}
\def \muo {\mu}

\title[Stratorotational instability in Taylor-Couette flow]{Stratorotational instability in Taylor-Couette flow heated from above}

\author[M. Gellert and G. R\"udiger]%
  {M.\ns G\ls E\ls L\ls L\ls E\ls R\ls T\ns and\ns G.\ns R\ls \"U\ls D\ls I\ls G\ls E\ls R}

\affiliation{Astrophysikalisches Institut Potsdam, An der Sternwarte 16, D-14467 Potsdam, Germany}

\pubyear{2009}
\volume{623}
\pagerange{375--385}
\date{2 September 2008 and in revised form 3 December 2008}
\begin{document}

\maketitle

\begin{abstract}
We investigate the instability and nonlinear saturation of temperature-stratified Taylor-Couette flows in a finite height cylindrical gap and 
calculate angular-momentum transport 
in the nonlinear regime. The model is based on an incompressible fluid in Boussinesq approximation with a positive axial temperature gradient 
applied. While both ingredients themselfs, the differential rotation as well as the stratification due to the temperature gradient, are stable, 
together the system becomes subject of the stratorotational instability and nonaxisymmetric flow pattern evolve. This flow configuration 
transports angular momentum outwards and will therefor be relevant for astrophysical applications.
The belonging coefficient of $\beta$-viscosity is of the order of unity if the results are adapted to the size of an accretion disc.
The strength of the stratification, the fluids Prandtl number and the boundary conditions applied in the simulations are well-suited too for
a laboratory experiment using water and a small temperature gradient around five Kelvin. With such a set-up the stratorotational instability and 
its angular momentum transport could be measured in an experiment.
\end{abstract}

\section{Introduction}
In recent years instabilities in stratified media became of higher interest. Especially in view of astrophysical objects the
inclusion of stratification is relevant. One simple model to study stratification effects is the classical Taylor-Couette (TC) system. \cite{thorpe_1968} found
a stabilizing effect of stratification only. In the context of stratorotational instability (SRI) stratification in TC flows is investigated numerically with 
fixed axial temperature gradient in a linear analysis studying the suppression of the onset of Taylor vortices by \cite{boub_1996}. With fixed density 
gradient \cite{mol_2001}, \cite{yav_2001} and \cite{shal_2005} show the onset of a linear instability and the growth of nonaxisymmetric modes. Experiments using 
artifically enlarged buoyancy due to salt concentration \cite[see][]{with_1974,boub_1995,lebars_2006} are in very good agreement with the linear results. 
\cite{caton_2000} use an 
artifical diffusivity in the continuity equation and show also results of linear stability analysis for the case of non-rotating outer cylinder. 
\cite{umur_2006} analyses SRI analytically, especially the influence of vertically changing buoyancy frequency, in the quasi-hydrostatic semi-geostrophic 
limit. The author also shows that SRI survives only in the presence of no-slip radial boundary conditions. In the context of accretion disks, 
\cite{dub_2005} figure out stability conditions and the influence of viscous dissipation and thermal diffusivity. In particular they find that the Prandtl number
dependence of the critical parameters is unimportant. But fully nonlinear three-dimensional
simulations do not exist. It is the aim of the first part of this paper to describe the characteristics of SRI in the nonlinear regime. The easiest way to do 
nonlinear simulations is the application of an axial temperature gradient, where the top of the TC system has higher temperature than the bottom.
The absence of a diffusivity in the continuity equation makes nonlinear simulations with explicit density gradient more demanding and favors a temperature
gradient. Further on this is interesting also from an astrophysical view point: accretion disks are heated from the central object at their top and 
bottom.  Thus the TC system heated from above could be seen as a simplified model for half of a disc. 
Of major interest for accretion or protostellar discs is the problem of angular momentum redistribution. Accretion only works if angular momentum is transported 
outwards effectively. Angular momentum transport of SRI is the second aspect of this work.

\section{Model}\label{sec_model}

The SRI is investigated in a model where stratification is due to a temperature gradient. At a first glance surprising this gradient is positive, i.e. the
cylindrical gap is heated from above. This configuration, opposite to the negative gradient in Rayleigh-B\'enard systems, is perfectly stable. The second 
ingredient in our model is differential rotation. The inner and outer cylinders rotate with different angular velocities. Both are chosen to be positive, i.e. 
exhibit the same direction of rotation. If the outer cylinder rotates sufficiently slow, the flow without temperature gradient evolves the well-known Taylor 
vortices for high enough Reynolds number. Beyond the Rayleigh line, when the outer cylinder rotates fast enough, the system is hydrodynamical 
stable. The combination of both stable parts can again drive the system unstable and generates a new instability, the SRI.

Stratification is measured by the Froude number 
\begin{equation}
 \Fr=\Om_{\rm in}/N,
\end{equation}
the ratio between angular velocity of the inner cylinder and stratification given by the buoyancy frequency 
\begin{equation}
 N^2 = \alpha g \frac{\p T}{\p z}
\end{equation}
Here $\alpha$ and $g$ are the coefficient of volume expansion and gravity respectively, $T$ is the temperature.

For the computations we fix the Froude number $\Fr$, the cylinder height and the Prandtl number. In \citet[]{rued_2008} it is shown that moderate
stratifications are most effective. This means that for too strong stratification the instability is suppressed as well as for too low values. For the latter 
in the limit $\Fr\rightarrow\infty$ the instability disappears in favor of the hydrodynamic stable configuration. Hence rotation and stratification of the same
order, i.e. $\Fr\approx1$, appear to be a good choice to excite the instability. \citet[]{rued_2008} find the optimum at $\Fr=1.4$, which is well-suited also 
for the nonlinear simulations presented here to get the smallest critical Reynolds numbers.

From an astrophysical point of view one of the most interesting questions arising is whether the SRI can transport angular momentum. If the normalized Reynolds 
stress $Q_{R\phi}/(R_{\rm in}\Om_{\rm in})^2$, where $R_{\rm in}$ and $\Om_{\rm in}$ are the inner radius and its angular velocity in cylindrical coordinates
$(R, \phi, z)$ respectively, defined by the velocity fluctuations correlation
\begin{equation}\label{eq_qrp}
Q_{R\phi} = \langle U'_R U'_\phi \rangle,
\end{equation}
turns out to be positive, angular momentum is transported outwards. This astrophysically important situation appears, as we show in~\S\,\ref{sect_angu}, 
always to be true in our simulations. Fluctuating quantities are defined as deviations from the azimuthally averaged mean field, 
i.e. $\f{U'}=\f{U}-\f{\langle U\rangle}$ with
\begin{equation}\label{eq_av}
 \langle \f{U}\rangle = \frac{1}{2\pi} \oint \f{U}\, {\rm d}\phi.
\end{equation}
Based on the idea, that angular momentum transport is mainly enhanced via an increase of viscosity, \citet[]{shak_1973} introduce in a 
parameterized model for a disc a coefficient
\begin{equation}
\alpha_{\rm SS} = \frac{Q_{R\phi}}{H^2 \Om_{\rm in}^2}
\end{equation}
with the density scale height $H$ of the disc. If the value of the $\alpha_{\rm SS}$ is of the order of unity, the angular momentum transport 
is considered to be very effective and important. Taking the normalized Reynolds stress $\beta=Q_{R\phi}/(R_{\rm in}\Om_{\rm in})^2$ for the SRI as 
an $\alpha$-like coefficient, this leads to
\begin{equation}
\alpha_{\rm SS} = \beta \left( \frac{R_{\rm in}}{H}\right)^2 \approx \beta \cdot 10^3
\end{equation}
with the so-called $\beta$ viscosity \cite[see][]{lyn_1974,hur_2001}. Here is assumed that the SRI survives if the cylindrical container becomes 
larger in radial direction, which is not yet clear.

\section{Equations and numerical treatment}\label{sect_num}
We use a code based on the nonisothermal hydrodynamic Fourier spectral element code described by \cite{fournier_2005} and \cite{gellert_2007}. With this 
approach we solve the 3D hydrodynamic equations in Boussinesq approximation
\begin{eqnarray}\label{eq_ns}
 \p_t \f{U}  + (\f{U}\bcdot\n)\f{U} &=& -\n p + \n^{\,2} \f{U} + \Gr\, {T}\, \hat{\bf{e}}_z, \\
 \p_t {T} + (\f{U}\bcdot\n) T &=& \frac{1}{\Pr} \n^{\,2} {T} \\
 \n\bcdot\f{U}  &=& 0
\end{eqnarray}
for an incompressible medium in a cylindrical annulus with inner radius $R_{\rm in}$ and outer radius $R_{\rm out}$.
Free parameters are the Grashof number 
\begin{equation}\label{eq_ra}
 \Gr = \frac{\Ra}{\Pr} = \frac{\alpha g \Delta T D^3}{\nu^2}
\end{equation}
and the Prandtl number $\Pr = \nu / \chi$.
Here $\nu$ is the viscosity of the fluid and $\chi$ its thermal conductivity. $D=R_{\rm out}-R_{\rm in}$ is the gap width. The Reynolds number 
appears not directly within the set of equations. It is defined, based on the inner cylinders angular velocity, as 
$\Re=\Om_{\rm in} R_{\rm in} D / \nu$ with the gap width $D$ (used as unit of length) and the angular velocity of the inner 
cylinder $\Om_{\rm in}$. Unit of velocity is $\nu/D$ and unit of time the viscous time $D^2/\nu$.

The solution is expanded in $M$ Fourier modes in the azimuthal direction. This gives rise to a collection of meridional problems, each of which is
solved using a Legendre spectral element method \cite[see e.g.][]{dev_2002}. Between $M=8$ and $M=32$ Fourier modes are used. The polynomial order 
is varied between $N=8$ and $N=16$ with three elements in radial direction. The number of elements in axial direction depends on the height of the cylinder, the
spatial resolution is the same as for the radial direction. 
With a semi-implicit approach consisting of second-order backward differentiation formula and third order Adams-Bashforth for the nonlinear forcing terms 
time stepping is done with second-order accuracy.

At the inner and outer walls no heat-flux conditions are applied. In axial direction solid end caps with fixed temperatures $T_0$ and $T_1$ delimitate
the cylindrical gap. The velocity on both the inner and outer cylinders is fixed to $\Om_{\rm in}$ and $\Om_{\rm out}$ respectively. On the end
caps stress-free conditions are applied to prevent Ekman circulation that would result from solid end caps rotating with
the angular velocity of the inner or outer cylinder. This restriction is not essential for the occurrence of the instability, if the aspect ratio $\Gamma$
is large enough. It simplifies the following analysis, because data along the full cylinder height can be included. 

As initial flow profile we use the typical Couette profile
\begin{equation}\label{eq_couette}
 \Om(R) = a + \frac{b}{R^2} 
\end{equation}
with
\begin{equation}
a = \frac{\muo - \heta^2}{1-\heta^2}\Om_{\rm in}, \qquad b = \frac{1-\muo}{1-\heta^2}R^2_{\rm in}\Om_{\rm in},
\end{equation}
the radius ratio $\heta=R_{\rm in}/R_{\rm out}$ and the ratio of angular velocities $\muo = \Om_{\rm out}/\Om_{\rm in}$.
As initial temperature distribution the linear static advection profile is applied.

Without stratification the critical Reynolds numbers for an infinite cylinder with $\heta=0.5$ is $\Re_{\rm crit}=68$ and $\Re_{\rm crit}=90$ for $\heta=0.78$. With
our simulations we find $\Re_{\rm crit}=69$ and $\Re_{\rm crit}=91$ respectively, which we decided to be in good agreement.

\section{Onset of instability}\label{sect_inst}

The SRI occurs as instability leading to nonaxisymmetric flow patterns. After a linear growth phase, nonlinear interactions of the unstable modes occur
resulting in a stable saturated state. The time the instability needs to evolve is around two viscous time scales $D^2/\nu$ or 120 system 
rotations, slightly depending on the Prandtl number. With $\Pr=1$ it needs slightly longer (150 rotations) and for $\Pr\geq7$ nearly 100 rotations. 
After saturation a stationary state is reached, where the flow pattern exhibits a drift compared to systems rotation.
The growth rate of the instability is rather slow compared to other instabilities to produce (MHD-)turbulence like the magnetorotational 
instability \cite[MRI, see][]{balb_1991} or Tayler instability \cite[TI, see][]{tayler_1957,rued_2007}, where it is of the order of a few rotations. 
For both instabilities magnetic effects play an essential role.  The SRI, even if slower, 
might be of comparable importance. As we will see in the following, it also leads to significant angular momentum transport. Further on stratification
can suppress MRI or TI and thereby it becomes the most efficient instability mechanism in environments with weak or very strong magnetic 
fields.  MRI and TI are suppressed by the strong magnetic field. Also in low-conducting environments like protostellar discs, where magnetic effects 
are unimportant, nonmagnetic instabilities could be of high relevance.

In our simulations we use two different gap sizes. The first one is a small gap with $\heta=0.78$ also utilised in \citet[]{shal_2005} and the second one is a 
wider gap with $\heta=0.5$.
Simulations are done with angular velocity ratios above the Rayleigh line. 
This means $\muo>0.25$ for $\heta=0.5$ and
$\muo>0.61$ for $\heta=0.78$. Two special values, $\muo=0.354$ for the wide gap and $\muo=0.69$ for the small gap, are most interesting. These are the 
quasi-keplerian profiles, i.e. inner and outer cylinders would rotate like planets after the keplerian law $\sim R^{-3/2}$, in between the profile differs slightly 
from a keplerian. After the Rayleigh stability criterion $\p_R (R^2\Om)^2>0$ the configurations in our simulations are always hydrodynamically stable
and the observed instability is due to the interplay between stratification and differential rotation.

\subsection{Wide gap}
The lowest critical Reynolds number for the onset of the instability is $\Re=285$ for $\heta=0.5$, thus a rather weak stratification supports the SRI most effective.
The most obvious difference between both gap sizes is the dominating unstable mode. For the wide gap it is always $m=1$ in the observed
parameter region \mbox{$\Re\leq1000$}, \mbox{$1\leq\Pr\leq10$} and \mbox{$0.7\leq\Fr\leq2$}. All higher modes appear with gradual lower energy, where the 
step from mode $m$ to $m+1$ is between 10 and 20\% of the higher mode. So the spectrum decreases very fast. 

\subsection{Small gap}
\begin{figure}
\includegraphics[width=0.31\textwidth]{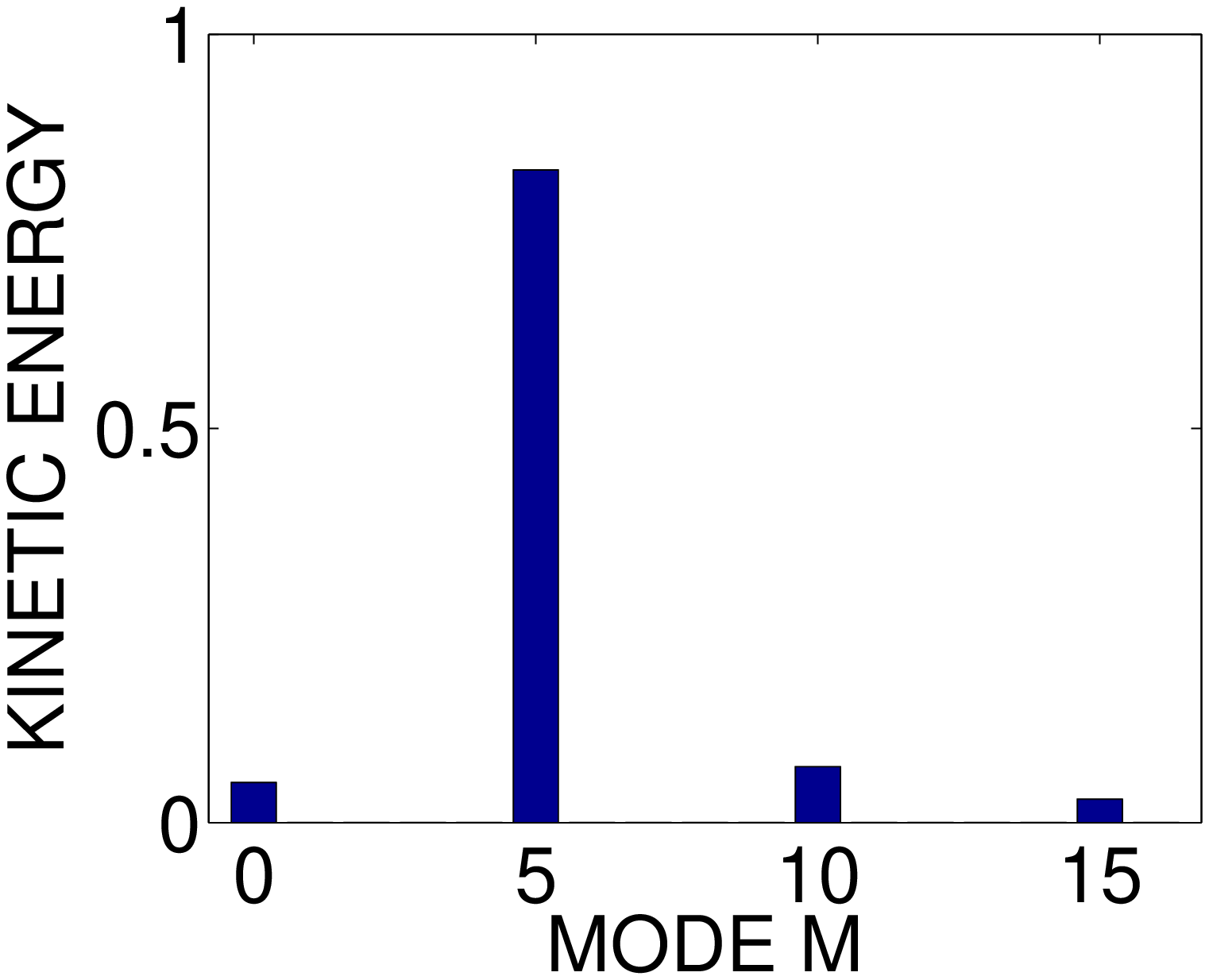}
\hfill
\includegraphics[width=0.31\textwidth]{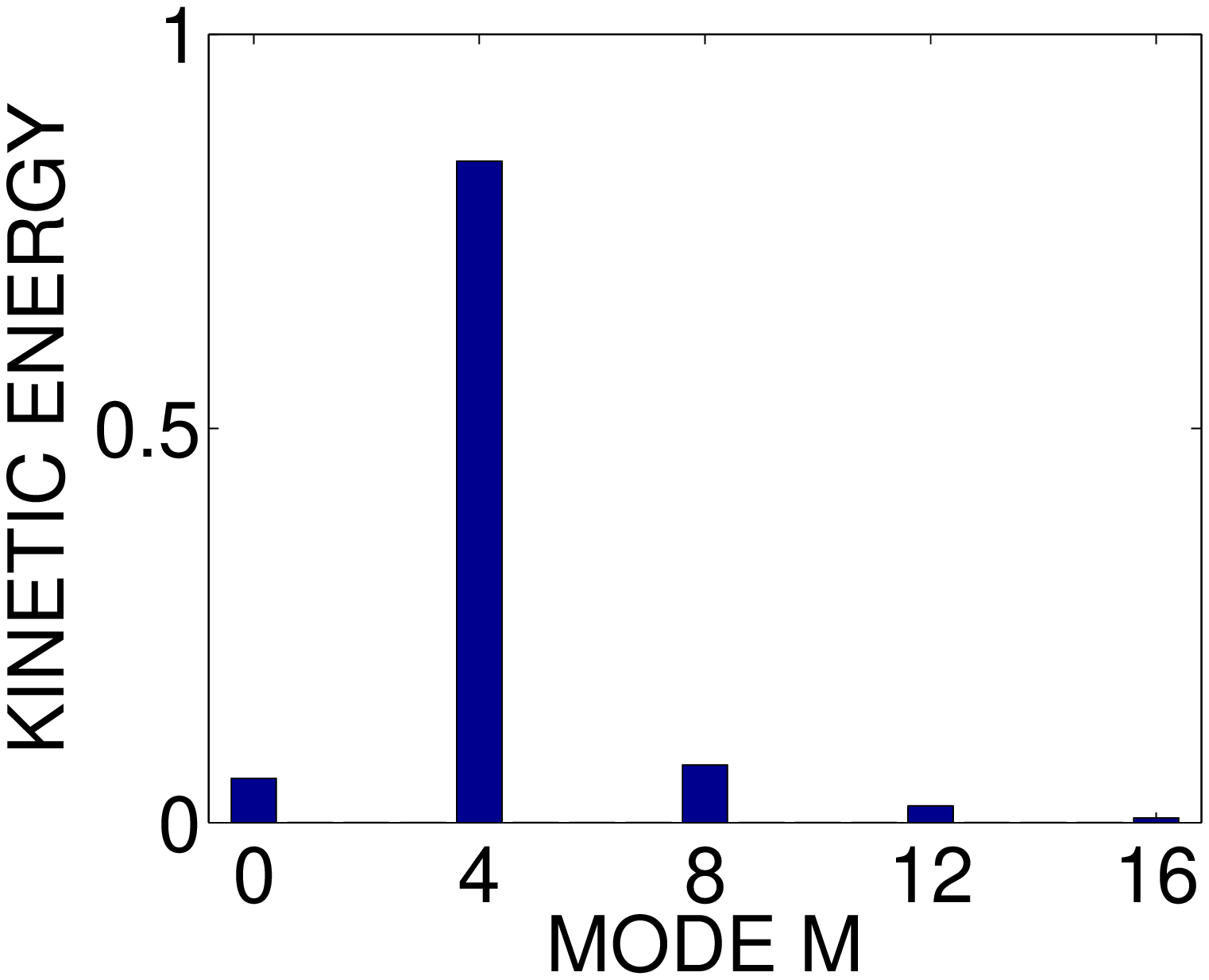}
\hfill
\includegraphics[width=0.31\textwidth]{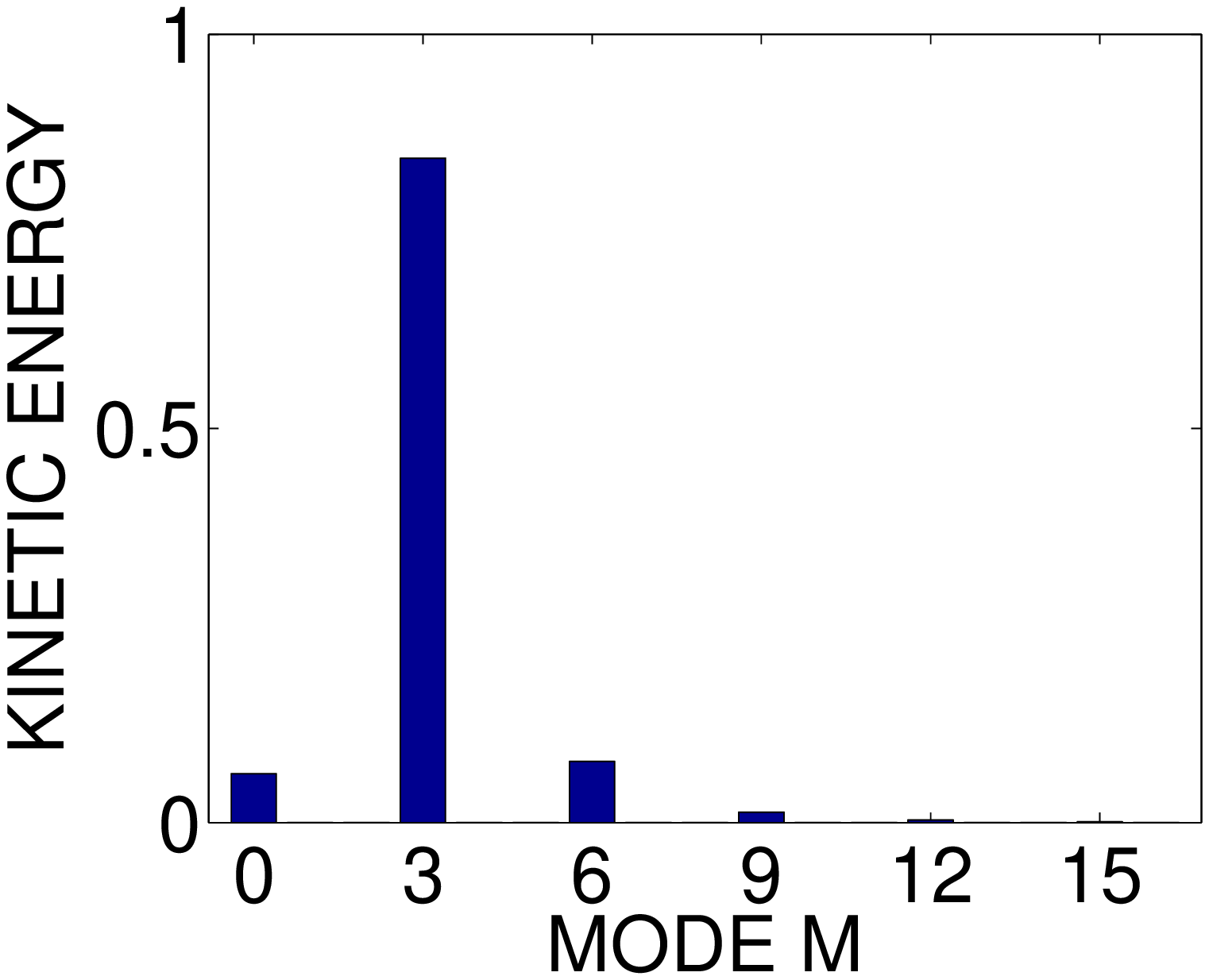}
\caption{Histograms of Fourier modes for $\Re=500, \Pr=7$ in the small gap $\heta=0.78$. The left plot is for $\Fr=0.7$, the middle for $\Fr=1$ and the right
         plot for $\Fr=1.4$. Shown is normalized total kinetic energy of the flow reduced by the driving due to the fixed cylinder rotation.}
\label{fig_hist}
\end{figure}

\begin{figure}
\includegraphics[width=0.31\textwidth]{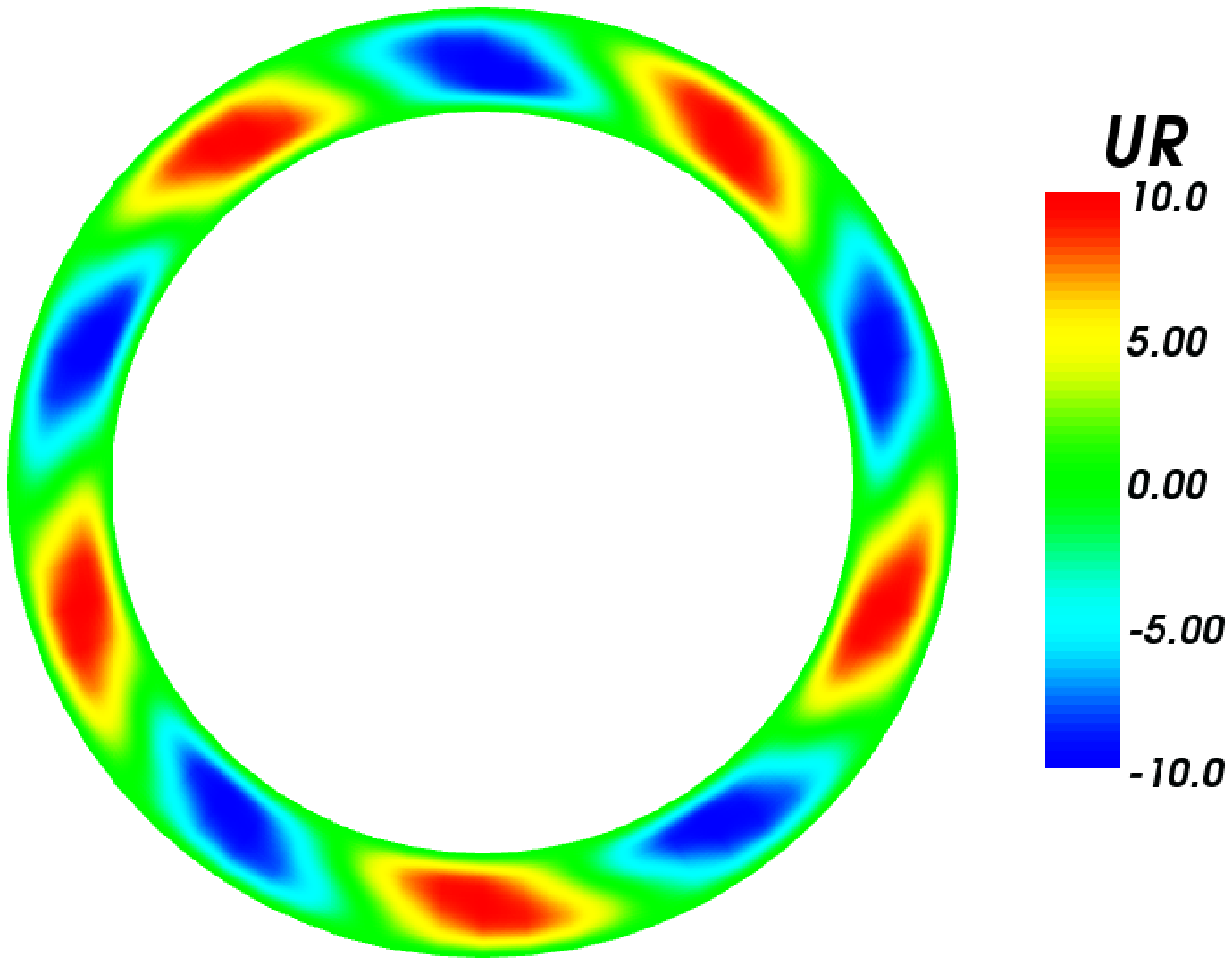}
\hfill
\includegraphics[width=0.31\textwidth]{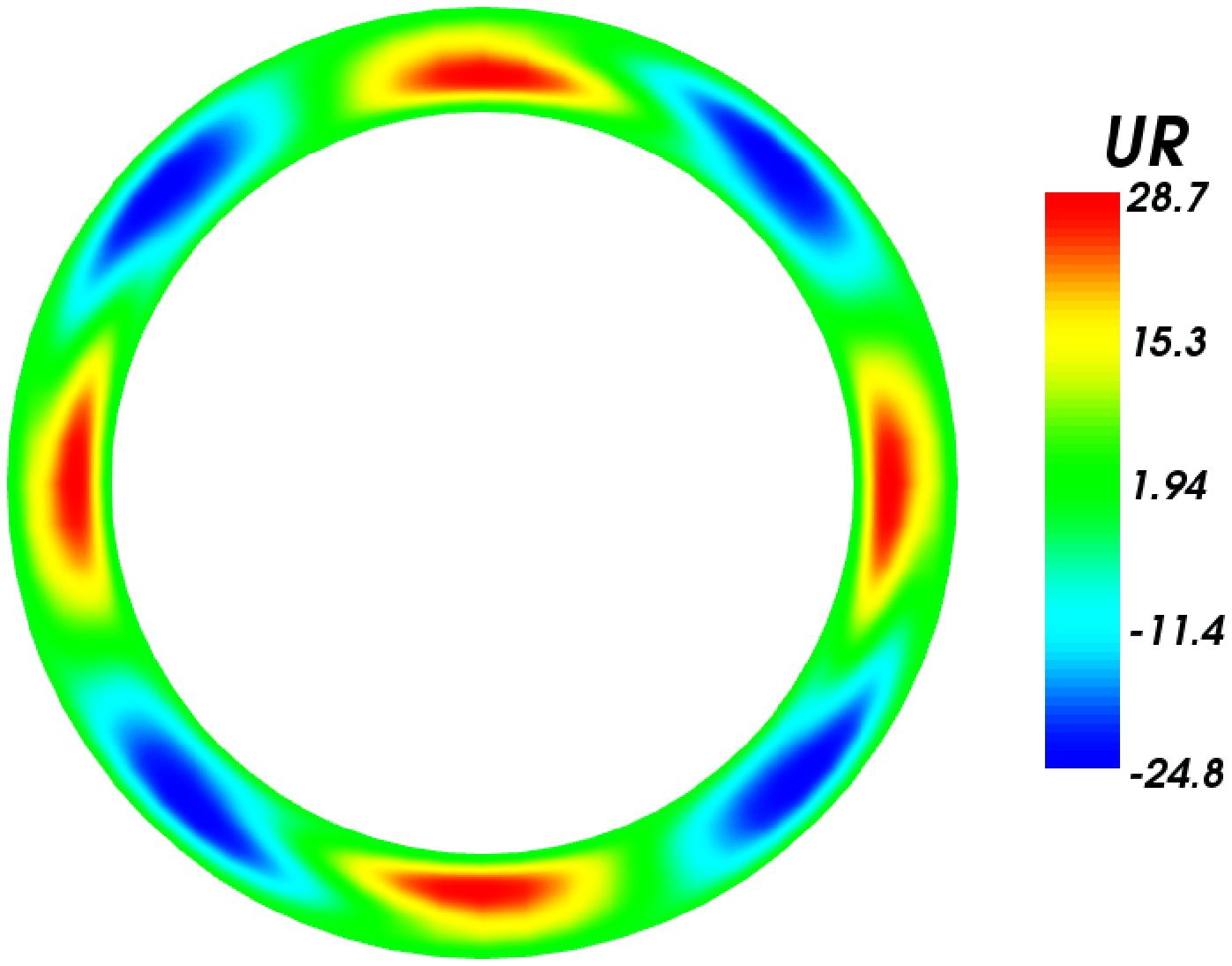}
\hfill
\includegraphics[width=0.31\textwidth]{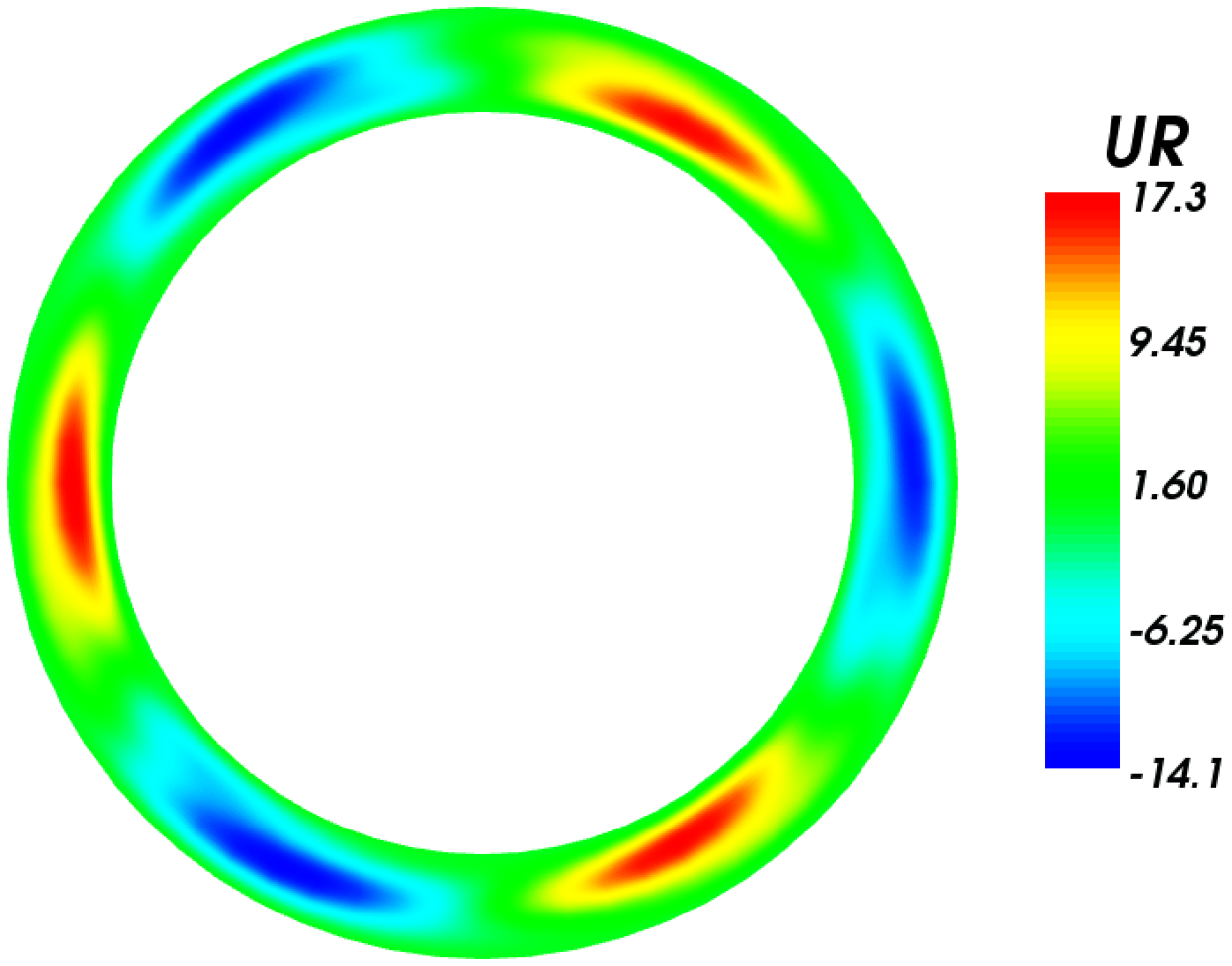}
\caption{Contour plots of radial velocity component $U_R$ in the midplane $z=\Gamma/2$ for $\Re=500, \Pr=7$ in the small gap $\heta=0.78$. The plots are for
         $\Fr=0.7$, $\Fr=1$ and $\Fr=1.4$ from left to right respectively. The dominating Fourier mode is determined by the strength of the stratification.}
\label{fig_cont}
\end{figure}

In the small gap case it is $m=3$, $m=4$ or $m=5$ that become linear unstable and grows exponentially. The lowest critical Reynolds number here is
$\Re=390$ for $\Fr=1.4$, again a weak stratification. When the unstable mode reaches a certain value,
nonlinear effects appear and a wide range of modes become excited. After reaching the saturation level, the energy of
the most unstable mode and of all harmonics stays constant, all other modes decay slowly on the thermal time scale.
The most unstable mode depends mainly on the stratification. For decreasing $\Fr=1.4,1.0,0.7$ the mode with the largest growth rate 
changes from $m=3$ to $m=4$ to $m=5$. And it is also this mode (and its harmonics) that dominates the saturated state (see figure \ref{fig_hist} and 
figure \ref{fig_cont}). For $\Fr=1.6$ it is $m=2$ that becomes the largest nonaxisymmetric mode and dominates the solution. For this rather low stratification one
needs already a larger Reynolds number of $\Re=1000$ to trigger the instability. So this solution is not directly comparable to all three other cases with $\Re=500$.
But the rule of decreasing $m$ with increase of $\Fr$ is still valid. 

\begin{figure}
\includegraphics[width=0.48\textwidth,height=0.18\textheight]{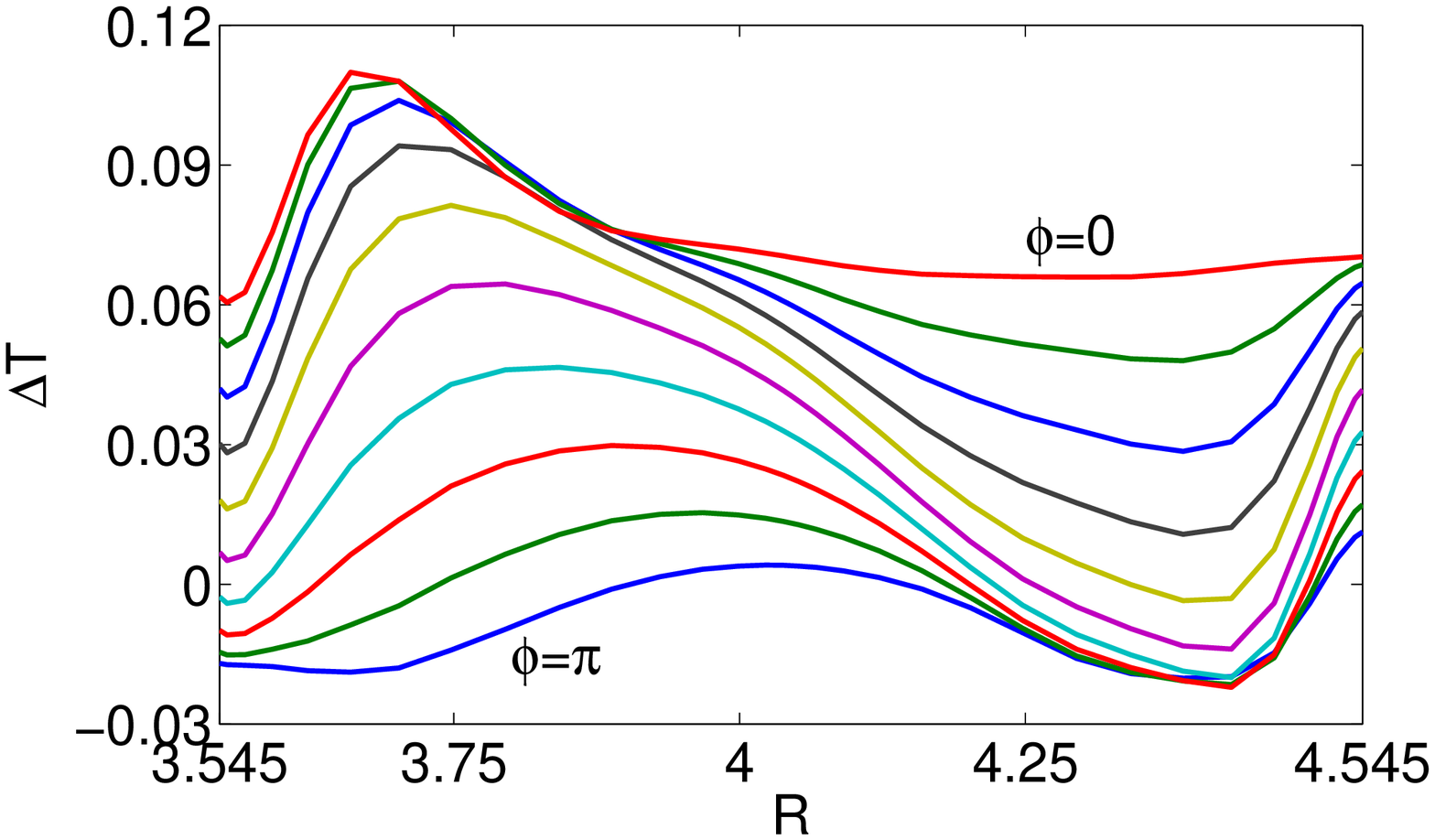}
\hfill
\includegraphics[width=0.48\textwidth,height=0.18\textheight]{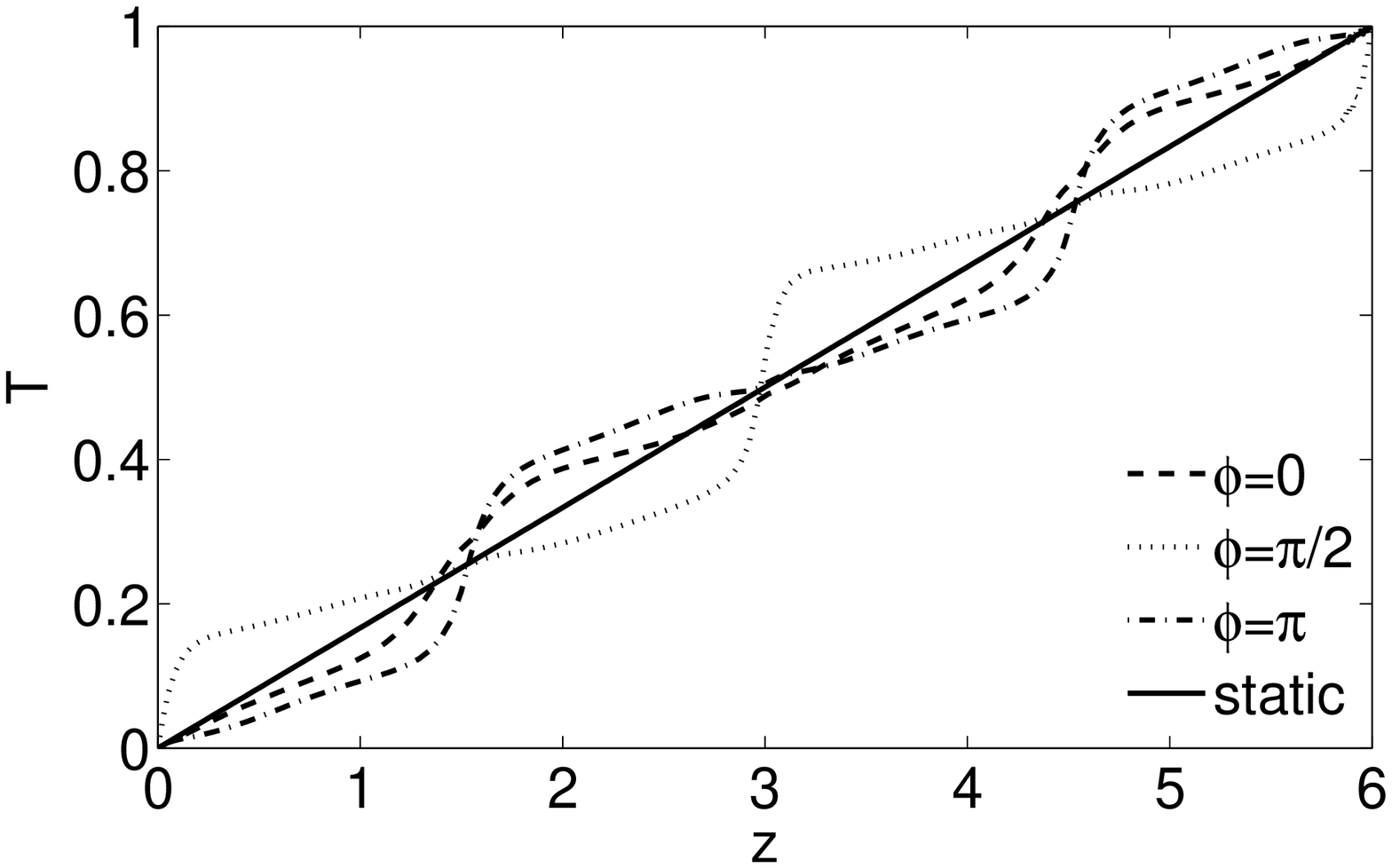}
\caption{Radial and axial temperature profiles for $\Re=700, \Pr=7, \Fr=1.4$ in the small gap for several values of $\phi$. Radial profiles are taken at half height
         and axial ones at the gap center. Deviations from the linear static profile are around 15\%.}
\label{fig_tempprof}
\end{figure}

The Prandtl number within the range $1\leq\Pr\leq10$ has no obvious
influence on the type of solution and the most unstable mode. Further on neither a variation of the Reynolds number (in the unstable regio) nor a 
variation of the steepness of the rotation profile $\muo$ has an obvious effect. This is different
to the results in \citet[]{shal_2005}, where $\muo$ and $\Re$ have a clear effect on the stability of the several modes. Either the lines of marginal stability for
the modes are so close to each other that a distinction in the nonlinear simulations is impossible, or the behavior changes remarkably when the stratification is
increased further. The lowest Froude number reachable in the nonlinear simulations is $\Fr=0.7$ and the linear analysis uses a value
of $\Fr=0.5$. Low $\Fr$ are more demanding because  $\Ra$ increases quadratically with $\Fr$ and also $\Re$ needs to be larger and limits the 
accessible parameter space. 
Another aspect is in agreement with the linear analysis. Already for $\muo=0.72$ we can not find
an instability for $\heta=0.78$ and $\Re\leq1000$, which is the maximum Reynolds number we can reach in our simulations. Thus, SRI feeds from an good balance 
between stratification and rotation with strong enough shear. 
The deviation from the linear static temperature profile is shown in figure \ref{fig_tempprof}. For a Reynolds number 1.5 times the critical one it differs significantly 
by around 15\% from the linear profile. This effect becomes larger with increasing \Re\ and vanishes near the onset of the stability.

\section{Angular momentum transport}\label{sect_angu}
As shown in~\S\,\ref{sect_inst}, the variety of flow pattern is larger for the small gap container and might be a good choice for a laboratory experiment.
Regarding angular momentum transport and its measurement a wider gap is favorable as we demonstrate in the following. With further aspects of an 
experimental realization is dealt in ~\S\,\ref{sect_exp}.

\subsection{Wide gap}
\begin{figure}
\begin{center}
\includegraphics[width=0.40\textwidth]{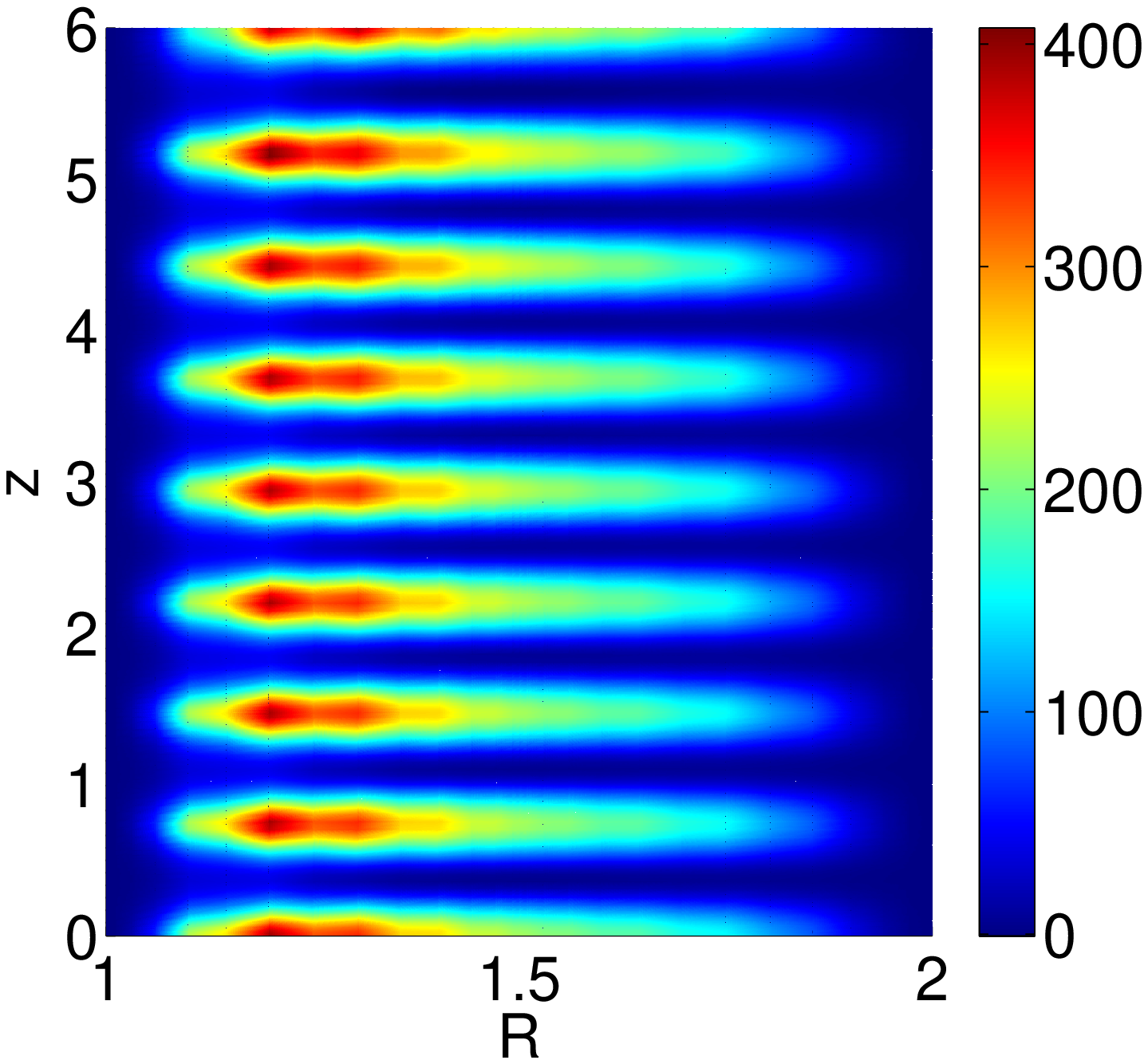}
\hfill
\includegraphics[width=0.56\textwidth]{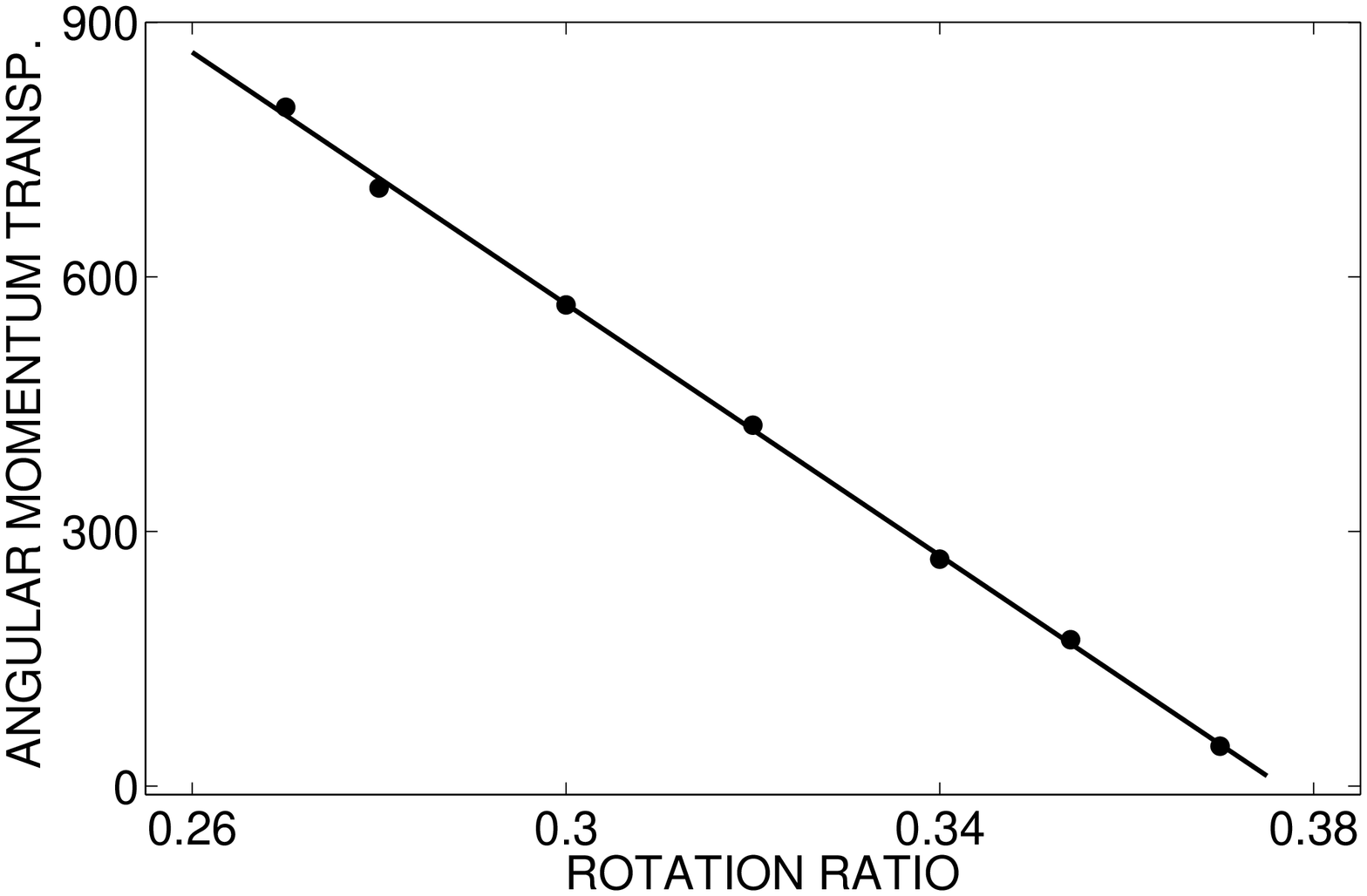}
\caption{Pattern of Reynolds stress $Q_{R\phi}=\langle U'_R U'_\phi \rangle$ (left) for $\muo=0.32$ and its linear dependence on shear (right). Angular momentum is
         transported outwards. Belonging parameters are $\Re=450, \Pr=1, \Fr=1$ and $\heta=0.5$.}
\label{fig_angu}
\end{center}
\end{figure}

By using equation (\ref{eq_qrp}) the Reynolds stress $Q_{R\phi}$ is calculated for a fixed set of parameters. A typical pattern in the $R-z-$plane is shown
in figure \ref{fig_angu} on the left. Angular momentum transport is always positive, i.e. directed outwards, with our configuration. The  axial wave number 
varies between 4 and 12 for $\Gamma=8$ depending mainly on the Froude number. Taking the value 
of $Q_{R\phi}$ in the gap center ($R=1.5$) averaged along $z$ gives the linear dependence on $\muo$ shown in figure \ref{fig_angu} on the right. Thus the 
angular momentum transport depends linearly on the shear.

\begin{figure}
\includegraphics[width=0.48\textwidth,height=0.22\textheight]{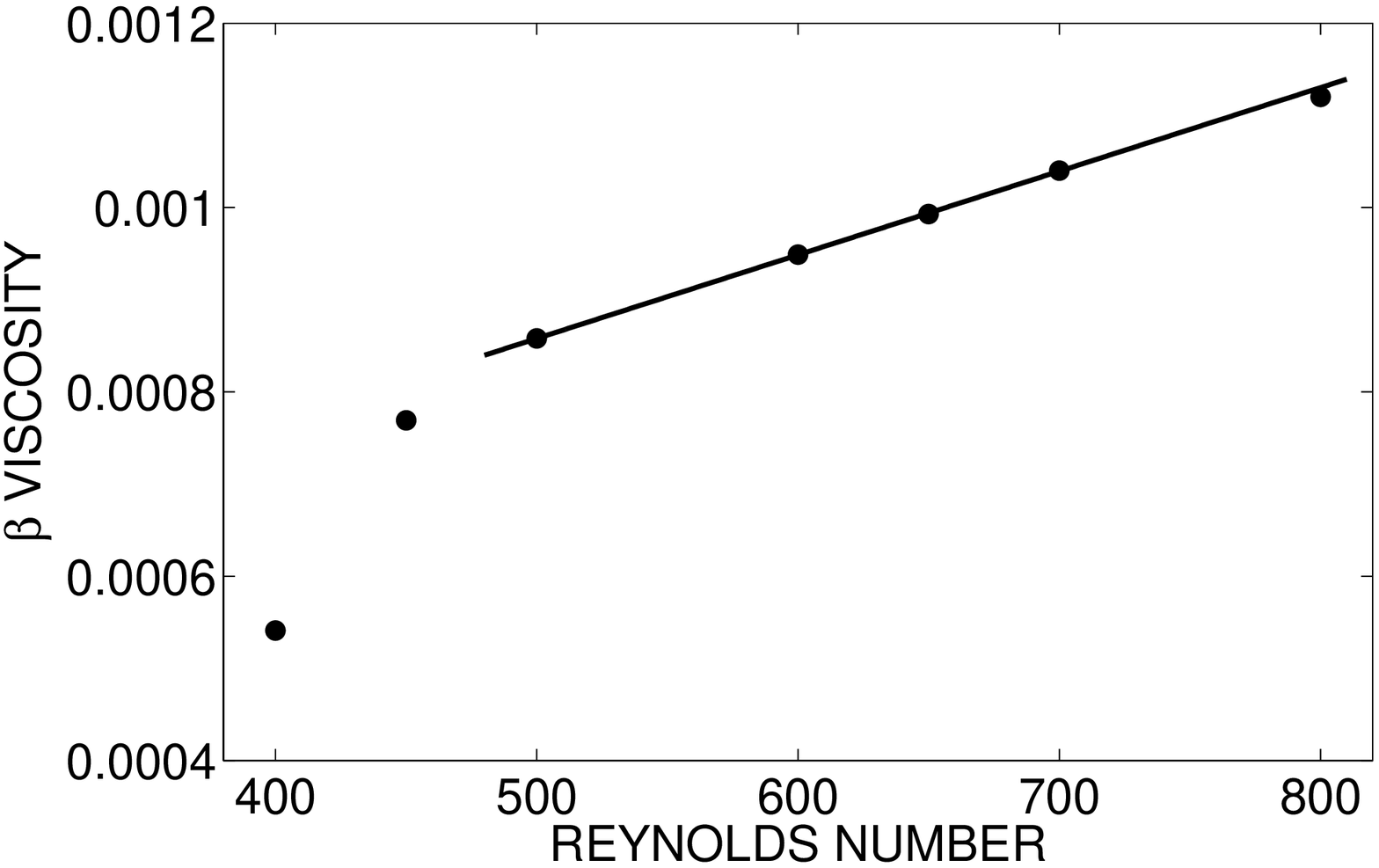}
\hfill
\includegraphics[width=0.48\textwidth,height=0.22\textheight]{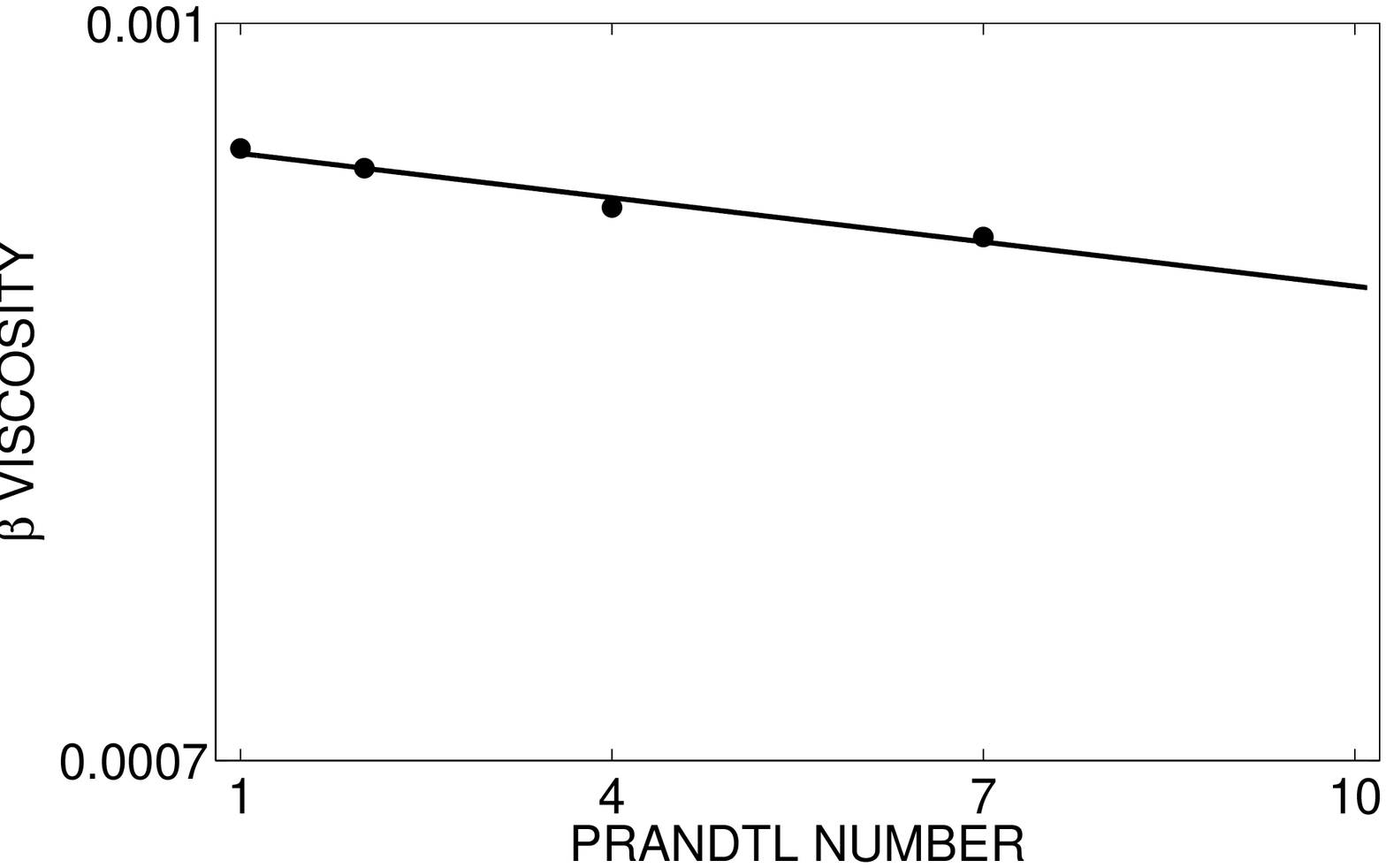}
\caption{For $\Fr=1.4$, $\muo=0.354$ and $\heta=0.5$: on the left angular momentum transport in terms of the $\beta$ viscosity at the gap 
         center ($R=1.5$) and averaged along $z$ for $\Pr=1$. For $\Re$ not in direct vicinity of the stability boundary, $\beta$ depends linear on $\Re$. 
         On the right side the weak dependence on the Prandtl number $\Pr$ is shown for $\Re=600$.}
\label{fig_angu_par}
\end{figure}

An increase of $\Fr$ leads to decreased angular momentum transport because maintenance of the instability needs a stronger driving flow for
increasing stratification. If there exists an maximum of $Q_{R\phi}$ for given $\Fr$ with increasing $\Re$ like it is the case for angular momentum 
transport due to the Tayler instability of a toroidal magnetic field \cite{gellert_2008}, is not easy to say. Within the accessible range of 
$\Re\leq1000$, $Q_{R\phi}$ increases linearly with $\Re$ and without indication for saturation (see figure \ref{fig_angu_par} left). A significant Prandtl 
number dependence can not be found (figure \ref{fig_angu_par} right). 

The nearly vanishing influence of $\Pr$ might be a sign for the fact, that it is not the convection-like temperature mixing that determines the SRI. This 
is only a secondary effect. The balance between centrifugal force and buoyancy is the crucial aspect. 
With $\chi\rightarrow0$ and constant $\nu$ the instability appears without qualitative change. The effect is only a slight decrease 
of $\beta$. On the other hand if both $\nu$ and $\chi$ together are decreased (constant $\Pr$ and increasing $\Re$), the $\beta$ viscosity grows rapidly
and linearly with $\Re$, which means a scaling $Q_{R\phi} \propto \Omega^3$.

\subsection{Small gap}
\begin{figure}
\includegraphics[width=0.48\textwidth,height=0.22\textheight]{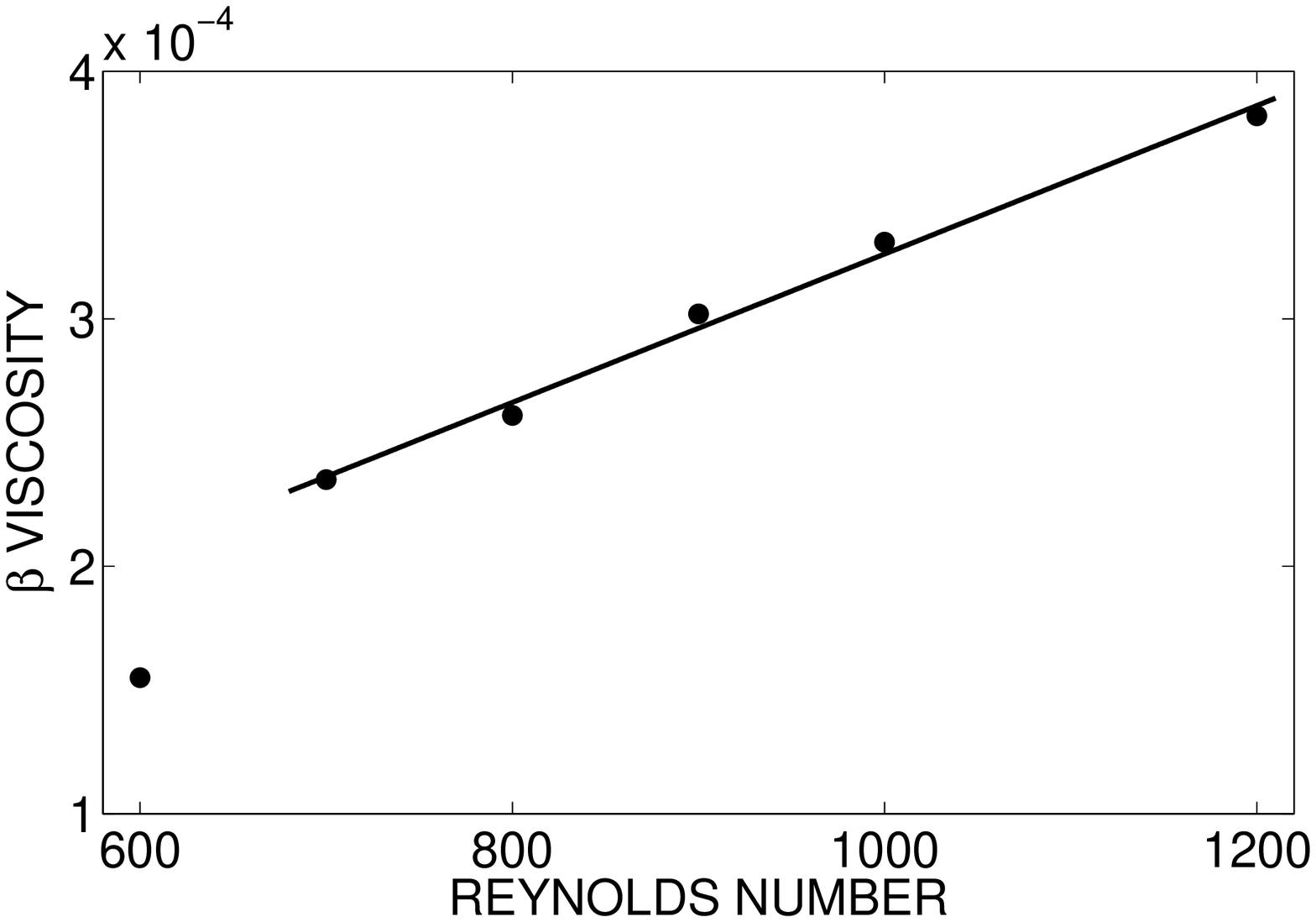}
\hfill
\includegraphics[width=0.48\textwidth,height=0.22\textheight]{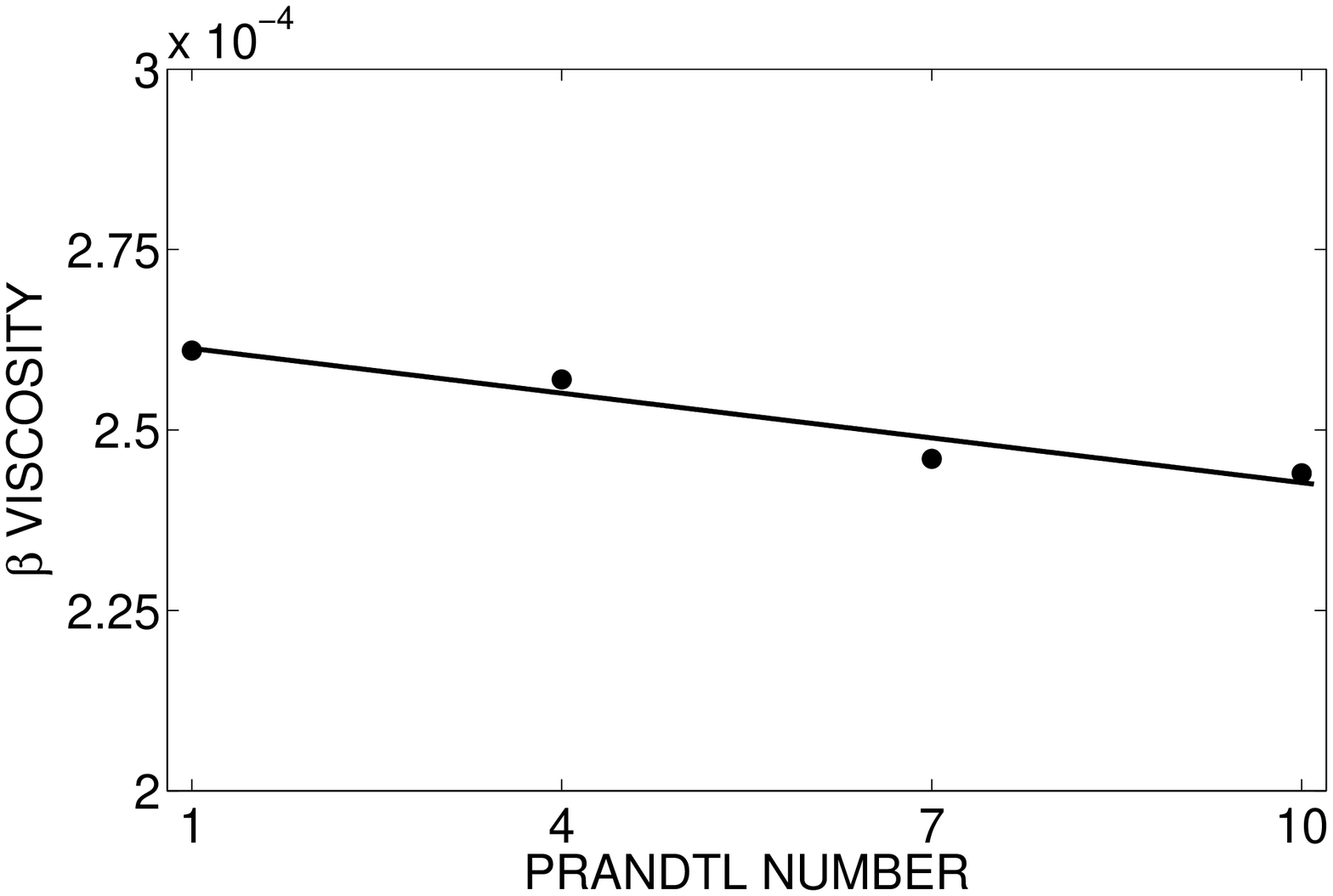}
\caption{Same as in Fig. \ref{fig_angu_par} but for the small gap with $\heta=0.78$ and $\Fr=1.4$. It is for $\Pr=1$ on the left and $\Re=800$ on the 
         right respectively.}
\label{fig_angu_par_small}
\end{figure}

Here the general behavior is the same as for the wide gap. The different nonaxisymmetric modes are not reflected in $\beta$. Compared with the small gap,
the $\beta$ viscosity is smaller by a factor of four but shows the same linear dependence on $\Re$ (see figure \ref{fig_angu_par_small}). The unimportant 
variation with the Prandtl 
number appears in the same way as for the wide gap for the observed Prandtl number range. This seems to be a very general behavior and is also reported 
for $\Pr\geq0.1$ by \cite{dub_2005}.

\section{Suggestion for a laboratory experiment}\label{sect_exp}

Experiments to study the SRI up to now use always a stratification accomplished by a salt solution. Test probes along the depth help to keep
the stratification linear as good as possible. Even if the unstable system would evolve deviations from the linear profile (and it would in the 
nonlinear regime as shown in figure \ref{fig_tempprof}), this is suppressed and the 
profile is forced to stay unchanged. This disadvantage can be avoided by using a temperature gradient to realise a strong enough stratification. Here 
the boundary conditions are well defined and it is a rather easy task to keep them constant in time. There exist a lot of experience, for instance from
Rayleigh-B\'enard experiments and other convective systems to work with temperature gradients in the lab. That is why such a set-up is promising. 

\begin{figure}
\includegraphics[width=0.32\textwidth]{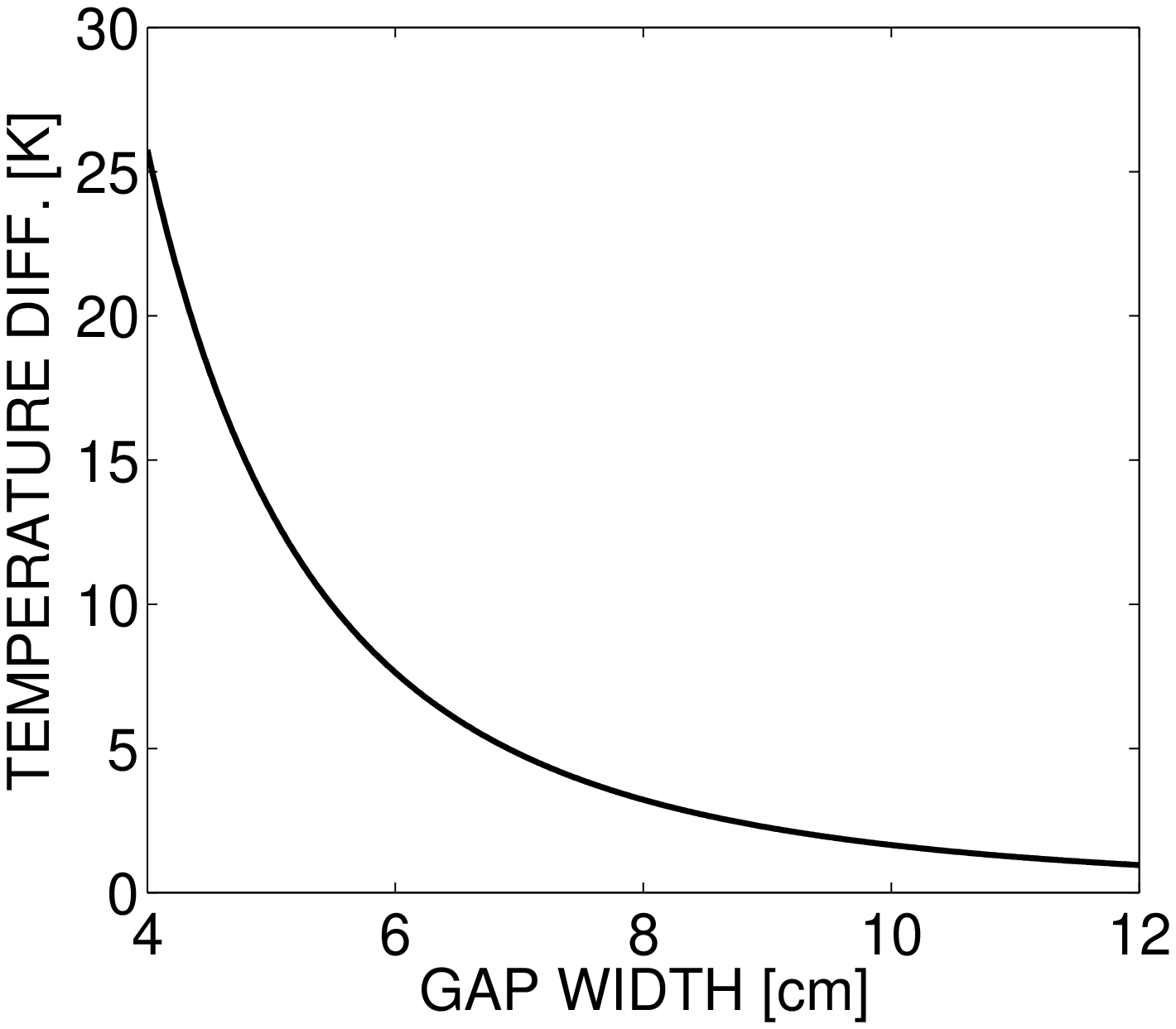}
\includegraphics[width=0.32\textwidth]{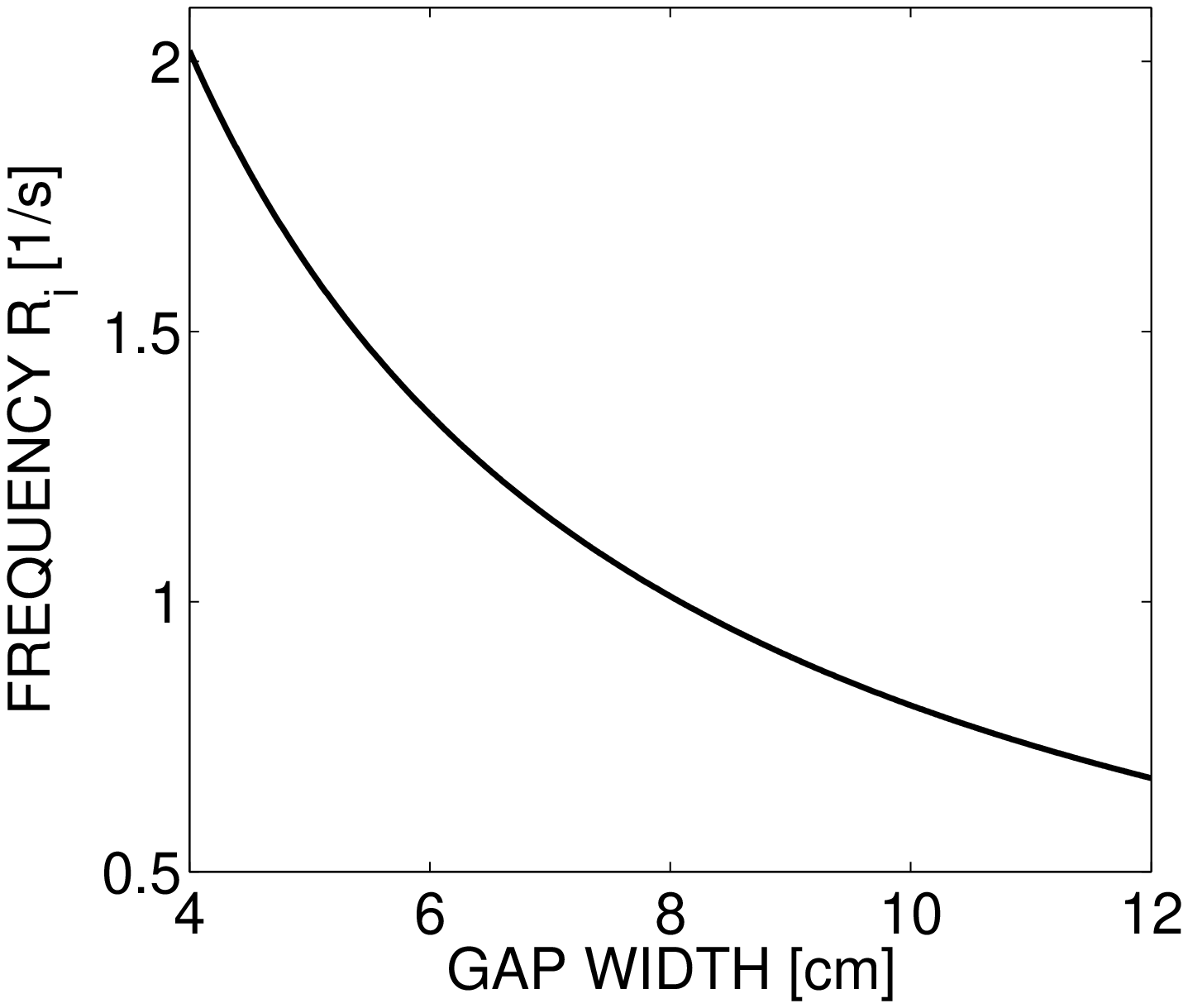}
\includegraphics[width=0.32\textwidth]{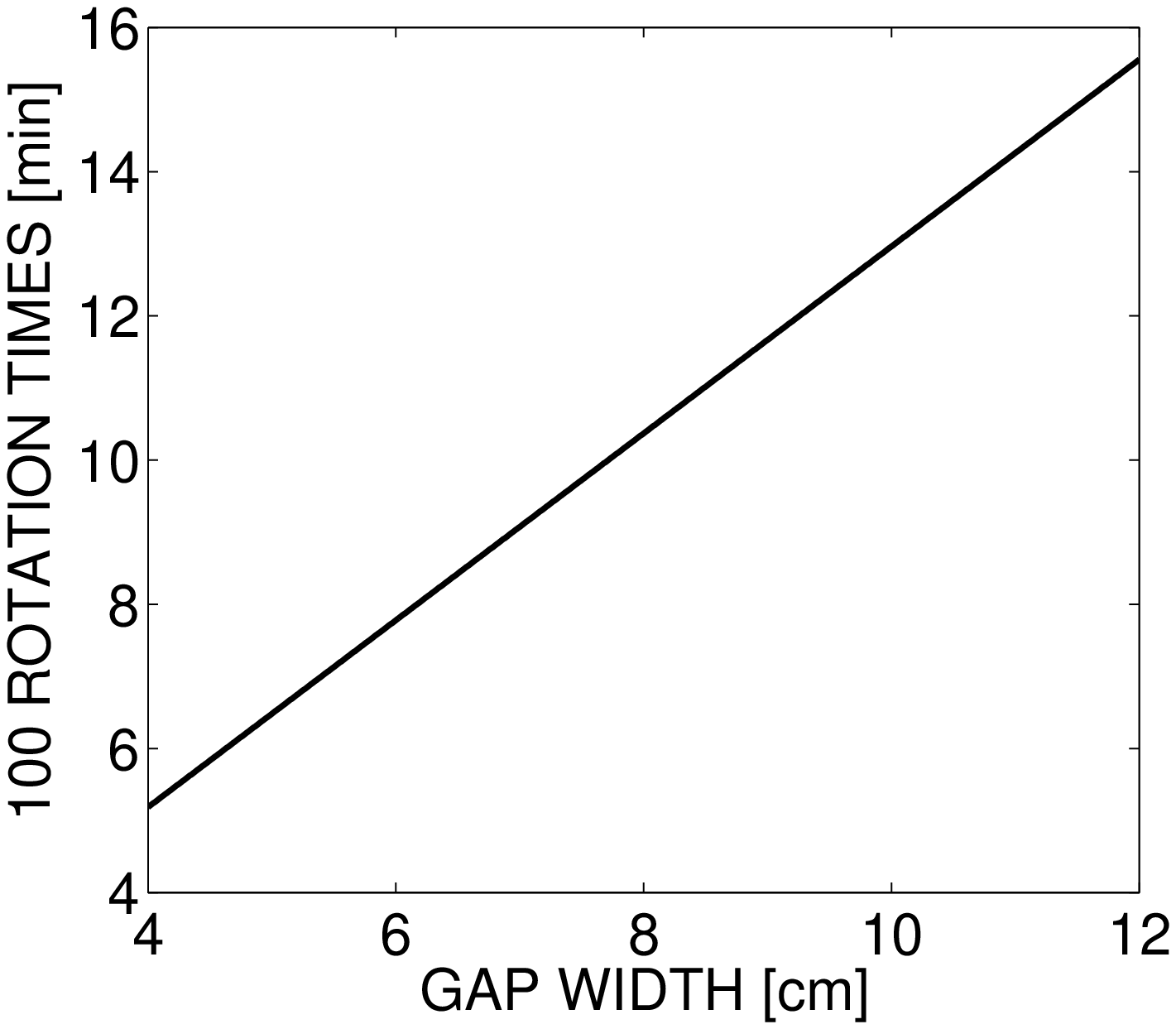}
\caption{Needed temperature difference between top and bottom (left), rotation frequency of inner cylinder (middle) and time to reach the saturation level (right)
         for $\heta=0.5, \Re=800, \Fr=1.4, \Gamma=10$ and $\Pr=7.16$ (water).}
\label{fig_dt05}
\end{figure}

We show in the following of what size the TC system should be to reach appropriate stratifications and rotation speeds for water as working fluid. The 
Prandtl number of water is $\Pr\approx7$.  For a possible set of parameters with $\heta=0.5, \Re=500, \Gamma=10$ and $\Fr=1.4$ one needs a Grashof number 
of $\Gr=1.3\cdot10^6$. This corresponds to a gap size of $6$ cm for a temperature difference of $5$ K or $9$ cm for $2$ K with water. Such dimensions for 
a TC system and the temperature difference are realizable in a laboratory. Figure \ref{fig_dt05} gives more details about dimensions, temperature differences
and rotation frequency of possible experiments with the same parameters but $\Re=800$.

The crucial point seems to be the choice of the aspect ratio of the cylinders. On one hand one would like to extend the system to a high aspect ratio to limit 
perturbing effects from the end caps. On the other hand increasing height linearly increases the required
temperature difference, because it is the derivative 
${\rm d}T/{\rm d}z$ that controls the instability and not the absolute value itself. Test simulations with solid end caps rotating with the angular 
velocity of the outer cylinder revealed that a smoothing of the pattern to axisymmetric structures near the top and bottom is inevitable, but it is 
the only effect if the aspect ratio is $\Gamma=8$ or higher. Between 20\% and 80\% of the height the instability appeared in the same way as without solid end 
caps. Thus an aspect ratio of $\Gamma=10$ might be an optimal choice for an experimental realisation.

Beneath the instability itself, also the transport of angular momentum should be possible to measure. A scenario to do so could be the following. The whole 
system is mounted on a rotation table rotating with the angular velocity of the inner cylinder. The outer cylinder has its own drive to realise the relative
angular velocity difference. The inner cylinder is connected to the table only by using thin wires. If the instability occurs an additional torque acts on
the inner cylinder and gives a measure for the $\beta$ viscosity.

\section{Discussion}
We have shown fully nonlinear simulations of the SRI with temperature stratification in a cylindrical annulus. The stratification is stable, as well as
the differentially rotating flow. Both together lead to unstable nonaxisymmetric modes. Depending on the gap width, these are the $m=1$ or $m=2/3/4/5$ modes
in the investigated parameter range of rather weak stratification with Froude numbers around $\Fr=1$. Weak stratification results in the lowest critical
Reynolds numbers for the onset of the SRI. On the other hand the instability is influenced by the Prandtl number of the flow only slightly. Especially
the dominating mode $m$ does not depend on $\Pr$, only on $\Fr$.
The time the SRI needs to evolve and reach a saturated state is of the order of 120 rotations or two times the viscous time scale. Thus the growth rate of 
the instability is rather slow compared 
to magnetorotational instability and Tayler instability, where it is of the order of ten rotations. For both instabilities magnetic effects play the essential 
role. The SRI, even if slower, might be of comparable importance, when magnetic effects are rather unimportant or if stratification suppresses MRI or TI. Thus it 
might become the most efficient instability mechanism in environments with weak or very strong magnetic fields or in low-conducting environments like protostellar 
discs.
We indeed find that SRI could be a mechanism able to transport angular momentum. The normalized angular momentum transport in terms of the
$\beta$ viscosity is of the order of $10^{-3}$, in terms of the $\alpha_{\rm SS}$ for a thin disk around unity. This comparison with a thin disk assumes
that SRI still occurs in such a flat disk, which is not possible to answer with our simulations at the moment. Nevertheless is the size of $\beta$ and its
linear growth with $\Re$ a sign of significant influence of the SRI for angular momentum transport which might dominate over magnetic effects to produce
turbulence in low-conducting environments. Beside astrophysical motivation angular momentum transport and turbulent transport coefficients are important 
also for technical applications.
In tradition of the idea of Couette to measure viscosity of a fluid, the presented experimental configuration could be used to measure the transport of angular 
momentum or the increase of viscosity in the laboratory. With water it needs a TC system with $\Gamma\approx 10$ and a gap width of \mbox{$6$ cm} to observe the SRI
with a temperature difference between top and bottom of \mbox{$5$ K}. For Reynolds numbers around $\Re=1000$ the time the instability needs to grow and to
saturate is around \mbox{$25$ min} or 120 rotations with \mbox{$0.5$ Hz}. All these conditions seem to be appropriate for an experiment. 

\begin{acknowledgments} 
The authors would like to thank Rainer Hollerbach for stimulating discussions.
\end{acknowledgments}


\end{document}